\documentclass{iaus}
\usepackage{graphicx}

\title[GMC Origins and Turbulent Motions] %% give here short title %%
{GMC Origins and Turbulent Motions in Spiral and Dwarf Galaxies}

\author[Bruce G. Elmegreen]   %% give here short author list %%
{Bruce G. Elmegreen}
\affiliation{IBM Research Division, T.J. Watson Research Center,
Yorktown Heights, NY USA \\ email: {\tt bge@us.ibm.com} }

\pubyear{2013}
\volume{292}  %% insert here IAU Symposium No.
\pagerange{1--4}
\jname{Molecular Gas, Dust, and Star Formation in Galaxies}
\editors{T. Wong and J. Ott}
\begin{document}

\maketitle

\begin{abstract}
CO clouds can be non-self-gravitating in high pressure environments, while
most should be strongly self-gravitating at low metallicities and ambient
pressures. In the LMC, which is HI-rich, GMC formation and destruction should
generally include molecule formation and destruction.  In M51, which is
CO-rich, GMCs grow by coalescence.  The Milky Way is between these two
situations. In all cases, large clouds form by accretion of gas and smaller
clouds independently of the presence of molecules. GMCs in the Milky Way are
analogous to dust lanes and spurs in other galaxies. The virial parameter
$\alpha$ usually decreases monotonically with increasing cloud mass in
surveys, which implies that small scale structure is formed by turbulence.
Hierarchies of sequences with decreasing $\alpha$ should be present in cloud
complexes from sub-solar masses up to the ambient Jeans mass
($10^7\;M_\odot$).

\keywords{stars: formation, ISM: clouds, galaxies: spiral, galaxies: dwarf}
\end{abstract}

\firstsection % if your document starts with a section,
\section{Introduction: the Molecular Transition}
Star-forming gas is usually traced by molecular line emission from CO. H$_2$
can be extensive without CO (e.g., \cite[Barriault et al. 2010]{barriault},
\cite[Lee et al. 2012]{lee}), although pure-H$_2$ is still apparently
confined to diffuse regions locally, not dense self-gravitating cores. For
this reason, CO line emission is fundamental for studying dense cloud
structures, and it is important to consider the different types of clouds
that emit CO. Such clouds need not be self-gravitating, for example. The
processes that assemble clouds are evident from observations, although
CO-cloud lifetimes are uncertain.

Local $^{13}$CO clouds have a minimum extinction in optical bands of around
1.5 magnitudes (\cite[Pineda et al. 2008]{pineda08}). This corresponds to a
mass column density of $\Sigma_{\rm CO}\sim 30\;M_\odot$ pc$^{-2}$ with solar
abundances, and to a self-gravitating pressure of $P_{\rm CO}\sim
(\pi/2)G\Sigma_{\rm CO}^2= 3\times10^4$ k$_{\rm B}$ K cm$^{-3}$. This is the
maximum boundary pressure for a marginally-CO cloud to be strongly
self-gravitating. Regions with higher pressures can have CO clouds that are
not strongly self-gravitating (``pressure-bound'' clouds). The threshold
pressure is higher than the average thermal pressure in the local ISM, so
local CO clouds are self-gravitating. Examples of CO clouds close to the
extinction threshold for CO-formation are the translucent clouds found by
\cite[Blitz et al. (1984)]{blitz} and studied more recently by
\cite[Barriault et al. (2010)]{barriault}. They contain CO in denser regions
but are generally pressure-bound.  CO in local diffuse clouds was measured in
absorption by \cite[Federman et al. (1980)]{federman}.

At lower metallicity, 1.5 magnitudes of visual extinction corresponds to a
higher mass column density, in inverse proportion to the dust-to-gas ratio.
Then the ambient pressure minimum for pressure-bound CO clouds is higher in
proportion to the square of the gas-to-dust ratio.  Clouds at the minimum
extinction for CO are therefore more likely to be strongly self-gravitating
at lower metallicity, as in dwarf galaxies.  The ambient pressure is also
lower in dwarfs, making any clouds that appear with CO even more
self-gravitating.

Molecular hydrogen appears in clouds that have much lower column densities
than CO-bearing clouds. In the solar neighborhood, the threshold for H$_2$
occurs at about $A_V\sim0.3$ mag (\cite[Spitzer \& Jenkins 1975]{spitzer}),
where the mass column density is $\Sigma_{\rm gas}=6 \;M_\odot$ pc$^{-2}$
(including He and heavy elements). This threshold is half the commonly
observed HI-H$_2$ transition in galaxies (\cite[Shaya \& Federman
1987]{shaya}, \cite[Bigiel et al. 2008]{bigiel}). The maximum pressure for
strongly self-gravitating, barely-$H_2$ clouds is $P_{\rm H2}\sim
(\pi/2)G\Sigma_{\rm H2}^2= 1200 k_{\rm B}$ K cm$^{-3}$, which is $\sim1/3$
the ambient thermal pressure in the solar neighborhood. Thus $H_2$-rich and
CO-poor clouds in the solar neighborhood are diffuse.

The actual appearance of CO in emission also depends on the density,
especially at low column density near the CO threshold where the CO lines are
optically thin. \cite[Krumholz (2011)]{krumholz11} shows that CO(1-0) is not
fully excited until the H$_2$ density reaches $\sim10^3$ cm$^{-3}$. A region
with the threshold $^{13}$CO column density of $A_V\sim1.5$ mag and an
average H$_2$ density of $10^3$ cm$^{-3}$ is only 0.45 pc thick. Such a
region, as a cubical volume, actually has a mass (6 $M_\odot$) that exceeds
the thermal Jeans mass (2.9 $M_\odot$) at 10K and the same density.  Note
that with this density, the internal pressure is comparable to the threshold
for CO given above and larger than the ambient thermal pressure, so fully
excited CO clouds are strongly self-gravitating at the local ambient thermal
pressure. A corollary to this statement is that local diffuse clouds close to
the CO-formation threshold should be only marginally excited in CO and
difficult to observe in emission. In higher pressure environments, such as
spiral arm dust lanes and inner regions of galaxies, diffuse clouds (i.e.,
those at the local pressure) close to the $^{13}$CO-formation threshold
should be well excited and more easily observed (\cite[Elmegreen 1993]{e93},
\cite[Shetty et al. 2012]{shetty}). For this reason, we expect high pressure
environments to contain observable diffuse CO, i.e., CO clouds that are not
strongly self-gravitating.

The transition to $H_2$ depends on both density and column density because
$H_2$ self-shields by absorption in saturated lines. Equilibrium is achieved
when the formation rate of $H_2$ on grains (a density-squared process
integrated over the path length of the shielding layer) equals the
destruction rate from external radiation. The threshold column density
therefore scales inversely with density for a given radiation field and
metallicity. \cite[Krumholz et al. (2008)]{krum08} write this as a dust
optical depth for the shielding layer $\tau=n\sigma_d z = \ln(1+\chi)$ where
$\chi\propto E/nZ$ for density $n$, dust cross section $\sigma_d$, shielding
length $z$, radiation field $E$, and metallicity $Z$, which enters into the
molecule formation rate on grain surfaces. For clouds shielded mostly by
H$_2$ self-absorption, $\chi<1$, $\ln(1+\chi)\sim\chi$ and the shielding
column density $nz$ is proportional to $1/n$ (times $E/Z$). For clouds
shielded by dust, $nz$ increases with $1/n$ logarithmically. Sufficiently
massive clouds in a given radiation field can be atomic and self-gravitating
because they have low densities in spite of their high column densities
(\cite[Elmegreen \& Elmegreen 1987]{e87}). The low densities result from high
velocity dispersions at a fixed ambient pressure. This is a manifestation of
Larson's scaling law between density and mass: at a given pressure for a
virialized cloud, the density scales inversely with the square root of the
mass. \cite[Pineda et al. (2008)]{pineda08} found a higher density threshold
for CO formation in higher dispersion clouds because of this opacity effect.

\section{Cloud Formation}

These relations between pressure, metallicity, self-gravity, and the presence
of molecules help us to understand the various ways in which molecular clouds
form.  Cloud formation or assembly by itself is the result of converging
flows that bring together ambient material and other clouds to make new,
larger clouds.  The flows can be initiated by self-gravity, turbulence,
expanding shells, spiral wave shocks, and other dynamical processes. When the
ISM metallicity or pressure are low, as in the LMC, the ISM is mostly atomic
and molecular clouds form by dynamical flows accompanied by a transition from
atoms to molecules. The molecules are in the self-gravitating parts of the
cloud. When star formation tears these clouds apart, they mostly revert back
into atomic gas. The result is a low molecular fraction and spotty CO
distribution, as observed in the LMC (\cite[Wong et al. 2011]{wong11}). When
the metallicity or pressure are high, as in M51 (\cite[Koda et al.
2009]{koda09}), or the radiation field is very low, then the ISM is mostly
molecular. Even a high fraction of diffuse clouds could be molecular, and the
most massive self-gravitating clouds (at the ambient Jeans mass of
$\sim10^7\;M_\odot$) can be molecular too. These are the giant molecular
associations in M51 (\cite[Vogel et al. 1988]{vogel}), which appear in M51's
spiral arms like similar-mass atomic clouds in the Milky Way (\cite[Elmegreen
\& Elmegreen 1987]{e87}). Cloud formation in a molecule-rich ISM is still by
dynamical flows, but the flows themselves are molecular. Then we can speak of
GMC formation as a process of GMC collisions, although physically is it
probably the same as in an atom-rich medium. In both cases, a clumpy,
turbulent medium converges to make giant clouds and strongly self-gravitating
cloud complexes.

The state of molecular gas in the Milky Way is somewhat between the extremes
of the LMC and M51: it is largely atomic near the solar radius and beyond,
and half-way molecular in the inner disk. There is apparently a transition
from mostly atomic gas in the interarm regions and some parts of spiral arms
to mostly molecular in the densest parts of the arms, which are presumably
spiral shocks, implying an HI-CO transition during cloud build-up followed by
some CO destruction as gas flows out of the arms. Because of the relatively
high spiral arm pressure, much of the CO in the spiral shocks can be in the
form of diffuse clouds. Most of the Milky Way's GMCs are in the cores of
giant HI clouds at the ambient Jeans-mass ($10^7\;M_\odot$) (\cite[Grabelsky
et al. 1987]{grabelsky}). The average densities of these clouds are low
($\sim9$ cm$^{-3}$; \cite[Elmegreen \& Elmegreen 1987]{e87}), and they should
be disrupted by shear and tidal forces when they emerge from the arms.

The association of massive Milky Way CO clouds with spiral arm shocks is
apparent in the distribution of CO from \cite[Dame et al. (2001)]{dame}
combined with kinematic models of the spiral arm gas flow (\cite[Bissantz et
al. 2003]{bissantz}). The concentration of CO in the arms is apparent in the
face-on view presented by \cite[Englmaier et al. (2011)]{englmaier}. These CO
clouds would appear as dust lanes in other galaxies.

\section{The Onset of Star Formation in GMCs}

Surveys of molecular clouds usually show a lack of self-gravity on small
scales, with increasing self-gravity for more massive clouds. This trend is
usually evident in a plot of the dimensionless virial parameter $\alpha$
($=M_{\rm virial}/M$) versus the cloud mass.  $\alpha$ is high for
non-self-gravitating clouds. In surveys, it is usually high for low mass
clouds and decreases with increasing mass (\cite[Dawson et al. 2008]{dawson},
\cite[Lada et al. 2008]{lada08}, \cite[Schlingman et al. 2011]{schlingman},
\cite[Barnes et al. 2011]{barnes11}, \cite[Belloche et al. 2011]{belloche},
\cite[Giannini et al. 2012]{giannini}).

The implication of this trend is that small clouds in homogeneous surveys
form by non-gravitational processes, which are most likely dominated by
ram-pressure driven convergence in a supersonically turbulent medium. Only
the largest scales are self-gravitating.

\cite[Hirota et al. (2011)]{hirota} show the change in $\alpha$ for GMCs that
enter a spiral arm in IC 342. Just before entering, the CO clouds have high
$\alpha$ and no star formation. Inside the arm, cloud velocity dispersions
and $\alpha$'s drop, and star formation begins. Spiral arm compression
appears to enhance turbulent dissipation and self-gravity in the clouds.

In a more comprehensive survey, which includes various molecules with
different density sensitivities, $\alpha$ would not necessarily show a
monotonic change with mass.  Most likely, clouds are a nested hierarchy of
self-gravitating and non-self-gravitating structures, with either occurring
inside the other. For example, self-gravity in the ambient ISM can make a
Jeans-mass cloud at $10^7\;M\odot$, and this cloud can fragment by turbulence
compression (producing an $\alpha$ sequence). The massive fragments then
become more self-gravitating after energy dissipation, forming GMCs with
locally low $\alpha$. These GMCs simultaneously fragment more by turbulence,
producing another $\alpha$ sequence inside of them on smaller scales.
Turbulent fragmentation stops when the sonic scale is reached. If the sonic
scale, which is initially high-$\alpha$, becomes self-gravitating, perhaps
after mass accretion, then the associated structure can form a star.  All of
this happens in a dynamically collapsing, turbulent region, so $\alpha$ is
likely to contain shearing and convergent or divergent motions.

We note the important difference between the gravitational time
($[G\rho]^{-1/2}$) and the crossing time ($R/\Delta V$). The ratio of these
two times is the square root of $\alpha$. For small regions inside a
virialized cloud, where $\alpha$ is large, the turbulent crossing time is
shorter than the gravitational time and turbulent motions dominate. To form a
star, such a region has to lower its $\alpha$ by energy dissipation and mass
accretion before the ambient flow disrupts it. If the small region is
non-isotropic, like a filament, then it can have effectively low $\alpha$ in
some directions and high $\alpha$ in other directions.  Filaments form stars
by accretion along the axis on a gravitational time, which is longer than the
crossing time on the minor axis.


\begin{thebibliography}{}

\bibitem[Barnes et al. (2011)]{barnes11} {Barnes, P. J., et al.} 2011,
    \textit{ApJS},
    196, 12

\bibitem[Barriault et al. (2010)]{barriault} {Barriault, L., et al.} 2010,
    \textit{MNRAS}, 406, 2713

\bibitem[Belloche et al. (2011)]{belloche} {Belloche, A., et al.}
    2011, \textit{A\&A}, 535, 2

\bibitem[Bigiel et al. (2008)]{bigiel} {Bigiel, F., et al.} 2008,
    \textit{AJ}, 136, 2846

\bibitem[Bissantz et al. (2003)]{bissantz} {Bissantz, N., Englmaier,
    P., Gerhard, O.} 2003, \textit{MNRAS}, 340, 949

\bibitem[Blitz et al. (1984)]{blitz} {Blitz, L., Magnani, L. \& Mundy, L.}
    1984, \textit{ApJ}, 282, L9

\bibitem[Dame et al. (2001)]{dame} {Dame, T. M., Hartmann, D., \& Thaddeus,
    P.} 2001, \textit{ApJ}, 547, 792

\bibitem[Dawson et al. (2008)]{dawson} {Dawson, J. R., et al.} 2008,
    \textit{MNRAS}, 387, 31

\bibitem[Elmegreen (1993)]{e93} {Elmegreen, B. G.} 1993, \textit{ApJ}, 411,
    170

\bibitem[Elmegreen \& Elmegreen (1987)]{e87} {Elmegreen, B. G. \&
    Elmegreen, D. M.} 1987, \textit{AJ}, 320, 182

\bibitem[Englmaier et al. (2011)]{englmaier} {Englmaier, P., Pohl, M., \&
    Bissantz, N.} 2011, \textit{MmSAI}, 18, 199

\bibitem[Federman et al. (1980)]{federman} {Federman, S. R., Glassgold, A.
    E., Jenkins, E. B., \& Shaya, E. J.} 1980, \textit{ApJ}, 242, 545

\bibitem[Giannini et al. (2012)]{giannini} {Giannini, T., et al.} 2012,
    \textit{A\&A}, 539, 156

\bibitem[Grabelsky, et al. (1987)]{grabelsky} {Grabelsky, D. A., Cohen, R.
    S., Bronfman, L., Thaddeus, P., \& May, J.} 1987, \textit{ApJ}, 315, 122

\bibitem[Hirota et al. (2011)]{hirota} {Hirota, A., Kuno, N., Sato,
    N., Nakanishi, H., Tosaki, T., \& Sorai, K.} 2011, \textit{ApJ}, 737, 40

\bibitem[Koda et al. (2009)]{koda09} {Koda, J., et al.} 2009, \textit{ApJ},
    700, L132

\bibitem[Krumholz et al. (2008)]{krum08} {Krumholz, M. R., McKee, C. F. \&
    Tumlinson, J.} 2008, \textit{ApJ}, 689, 865

\bibitem[Krumholz (2011)]{krumholz11} {Krumholz, M. R.} 2011, in XVth Special
    Course of the National Observatory of Rio de Janeiro, arXiv:1101.5172

\bibitem[Lada et al. (2008)]{lada08} {Lada, C. J., Muench, A. A.,
    Rathborne, J., Alves, J. F., Lombardi, M.} 2011, \textit{ApJ}, 672, 410

\bibitem[Lee et al. (2012)]{lee} {Lee, M.-Y., et al.} 2012, \textit{ApJ},
    748, 75

\bibitem[Pineda et al. (2008)]{pineda08} {Pineda, J. E., Caselli, P., \&
    Goodman, A. A.} 2008, \textit{ApJ}, 679, 481

\bibitem[Schlingman et al. (2011)]{schlingman} {Schlingman, W. M., et al.}
    2011, \textit{ApJS}, 195, 14

\bibitem[Shaya \& Federman (1987)]{shaya} {Shaya, E. J., \& Federman, S. R.}
    1987, \textit{ApJ}, 319, 76

\bibitem[Shetty et al. (2012)]{shetty} {Shetty, R., Kelly, B. C., \& Bigiel,
    F.} 2012, preprint

\bibitem[Spitzer \& Jenkins (1975)]{spitzer} {Spitzer, L., Jr., \& Jenkins,
    E. B.} 1975, \textit{ARA\&A}, 13, 133

\bibitem[Vogel et al. (1988)]{vogel} {Vogel, S. N., Kulkarni, S. R., \&
    Scoville, N. Z.} 1988, {\textit Nature}, 334, 402

\bibitem[Wong et al. (2011)]{wong11} {Wong, T., et al.} 2011, \textit{ApJS},
    197, 16

\end{thebibliography}
\end{document}